\documentstyle[11pt]{article}

\newcounter{itemlistc}
\newcounter{romanlistc}
\newcounter{alphlistc}
\newcounter{arabiclistc}

\begin{document}
\baselineskip=8truemm
\setcounter{page}{1}

\centerline{\normalsize\bf SO(10) GUT MODELS AND COSMOLOGY}
\vspace{0.6cm}
\centerline{\footnotesize GIAMPIERO ESPOSITO}
\baselineskip=13pt
\centerline{\footnotesize\it Istituto Nazionale di Fisica Nucleare,
Sezione di Napoli}
\baselineskip=12pt
\centerline{\footnotesize\it Mostra d'Oltremare Padiglione 20,
80125 Napoli, Italy}
\baselineskip=12pt
\centerline{\footnotesize\it Dipartimento di Scienze
Fisiche, Mostra d'Oltremare Padiglione 19, 80125 Napoli, Italy}
\centerline{\footnotesize E-mail: esposito@napoli.infn.it}
\vskip 1cm
\baselineskip=13pt

\leftskip=12mm
\rightskip=12mm

{\small
\noindent
\hskip 12mm {\bf Abstract.} $SO(10)$ grand unified models have an
intermediate symmetry group in between $SO(10)$ and 
$SU(3)_{C} \otimes SU(2)_{L} \otimes U(1)_{Y}$. Hence they lead to
a prediction for proton lifetime in agreement with the experimental
lower limit. This paper reviews the recent work on the tree-level
potential and the one-loop effective potential for such models,
with application to inflationary cosmology. The open problems are
the use of the most general form of tree-level potential for 
$SO(10)$ models in the reheating stage of the early universe, and the
analysis of non-local effects in the semiclassical field equations
for such models in Friedmann-Robertson-Walker backgrounds.}

\leftskip=0mm
\rightskip=0mm

\baselineskip=5truemm

\vskip 1cm
\noindent
\leftline {\bf 1. Introduction}
\vskip 1cm
\noindent
A key tool in modern particle physics is the spontaneous symmetry
breaking mechanism, which makes it possible to unify electroweak
and strong interactions in such a way that all mass terms for vector
mesons are generated without violating gauge invariance and
perturbative renormalization. Among the various possible choices for
a local symmetry group, models which rely on the $SO(10)$ group are
still receiving careful consideration in the literature. The main
motivations are as follows [1--3]:
\vskip 0.3cm
\noindent
(i) $SO(10)$ models make it possible to obtain masses of the
lepto-quarks mediating proton decay which are higher than the
ones found within the minimal $SU(5)$ model. Hence one recovers
agreement with the experimental lower limit of 
$9. \times 10^{32}$ years for proton decay.
\vskip 0.3cm
\noindent
(ii) All exceptional gauge groups contain $SO(10)$ as a subgroup.
\vskip 0.3cm
\noindent
(iii) Exact cancellation of chiral anomalies is obtained.
\vskip 0.3cm
\noindent
(iv) The only $SO(2n)$ group which contains a 16-dimensional
representation is indeed $SO(10)$. This makes it possible to
accommodate fermions and antifermions of each generation in a
single 16-dimensional complex representation,
while the introduction of exotic or mirror particles is avoided.
\vskip 0.3cm
\noindent
(v) On requiring that the GUT group should contain $SU(2)_{L}$ 
and $B-\alpha L$ as local symmetries, while no exotic particles
or mirror fermions should occur, one finds that $SO(10)$ is
unique and has $B-L$ as generator (hence $\alpha=1$).
\vskip 0.3cm
\noindent
(vi) $SO(10)$ models predict masses for $\tau$- and $\mu$-neutrinos
of the order of magnitude relevant for cosmology and solar-neutrino
astrophysics (including the effects of renormalization-group
equations [3]).

Section 2 outlines properties and symmetry-breaking pattern for
$SO(10)$ models. Sections 3 and 4 are devoted to the tree-level
potential and one-loop effective potential 
respectively. The one-loop
analysis is performed in a cosmological background, i.e. de Sitter
space. Results and open problems are presented in section 5.
\vskip 1cm
\leftline {\bf 2. Structure and symmetry breaking of
$SO(10)$ models}
\vskip 1cm
\noindent
The group $SO(10)$ consists of all $10 \times 10$ orthogonal
matrices with unit determinant, and with the usual product rules. 
It has 45 generators, say $T_{ij}$ ($i,j=0,1,...,9$), which obey
the commutation relations [4]
$$
\Bigr[T_{jk},T_{lm}\Bigr]=i \Bigr(\delta_{jl}T_{mk}
+\delta_{jm}T_{kl}+\delta_{kl}T_{jm}+\delta_{km}T_{lj}\Bigr) \; .
\eqno (2.1)
$$
Given the vector irreducible representation $\underline{10}$ 
of $SO(10)$, the action of the generators on 
$\varphi_{l} \in \underline{10}$ is given by 
$T_{jk}\varphi_{l} \equiv i \Bigr(\delta_{kl}\varphi_{j}
-\delta_{jl}\varphi_{k}\Bigr)$.

To break the symmetry spontaneously, one needs a Higgs field
belonging to one or more irreducible representations of the gauge
group $SO(10)$. In particular, one is interested (see section 4)
in the most general, renormalizable and conformally invariant
Higgs potential constructed by using only the 210-dimensional
irreducible representation. Such a representation has four
independent quartic invariants, i.e. the fourth power of the norm
of the Higgs field and three non-trivial invariants [4].
The tensor product of two $\underline{210}$
representations, jointly with symmetrization, yields the
fundamental Clebsch-Gordan decomposition [4]
$$
(\underline{210} \otimes \underline{210})_{\mbox {sym}}
=\underline{1} \oplus \underline{45} \oplus \underline{54}
\oplus \underline{210} \oplus \underline{770} \oplus
(\underline{1050} \oplus \overline{\underline{1050}})
\oplus \underline{4125} \oplus \underline{8910}
\oplus \underline{5940} \; ,
\eqno (2.2)
$$
where $\underline{45}$ denotes the 45-dimensional irreducible 
representation of $SO(10)$, and similarly for the others.

The symmetry breaking process of $SO(10)$ models consists of
the following three steps:
$$
SO(10) ~ {\mathrel{\mathop{\longrightarrow}^{M_{X}}}} G'
~ {\mathrel{\mathop{\longrightarrow}^{M_{R}}}}
SU(3)_{C} \otimes SU(2)_{L} \otimes U(1)_{Y} 
\longrightarrow SU(3)_{C} \otimes U(1)_{Q} \; .
\eqno (2.3)
$$
In the first stage, symmetry is broken at the scale $M_{X}$
of order $3.2 \times 10^{15}$ GeV, and one gets an intermediate
symmetry group $G'$ which may take the forms 
$\Omega_{1} \times D, \Omega_{1}, \Omega_{2} \times D,
\Omega_{2}$, where $\Omega_{1} \equiv SU(4)_{PS} \otimes SU(2)_{L}
\otimes SU(2)_{R}$, $\Omega_{2} \equiv SU(3)_{C} \otimes SU(2)_{L}
\otimes SU(2)_{R} \otimes U(1)_{B-L}$ and $D$ is the discrete
left-right interchanging symmetry [3, 4]. If $G'=\Omega_{1} \times D$,
the corresponding Higgs field $\Phi$ belongs to the 
54-dimensional irreducible representation of $SO(10)$. Otherwise,
$\Phi$ belongs to the 210-dimensional irreducible representation.
The second symmetry breaking occurs at the scale $M_{R}$ of order
$10^{11}$ GeV, by means of a scalar field $\varphi_{+} \in
\underline{126}$, and a scalar field $\varphi_{-} \in 
\overline{\underline{126}}$. Last, the symmetry breaking to
$SU(3)_{C} \otimes U(1)_{Q}$ involves a scalar field 
$\rho \in \underline{10}$.
\vskip 10cm
\leftline {\bf 3. Tree-level potential}
\vskip 1cm
\noindent
Following Ref. [3], we are here interested in the orbit structure of 
the tree-level potential for $SO(10)$ GUT models in flat space-time.
For this purpose, we denote by $\Phi,\varphi_{+},\varphi_{-}$ and
$\rho$ the generic elements of the $\underline{210}, \underline{126},
\overline{\underline{126}}$ and $\underline{10}$ representations,
respectively. Hence we can write the tree-level potential in the
form [3]
$$
V(\Phi,\varphi_{+},\varphi_{-},\rho)=V_{0}+V_{\Phi}(\Phi)
+V_{\varphi}(\varphi_{+},\varphi_{-})
+V_{\Phi,\varphi}(\Phi,\varphi_{+},\varphi_{-})
+V_{\Phi,\varphi,\rho}(\Phi,\varphi_{+},\varphi_{-},\rho) \; .
\eqno (3.1)
$$

The term $V_{0}$ only depends on the fields' norms and hence
cannot affect the direction of the potential minimum [5]. The
first symmetry breaking is instead realized by $V_{\Phi}(\Phi)$,
which breaks $SO(10)$ to the group $G'$ of section 2. The
second symmetry breaking is then realized by means of 
$\Bigr[V_{\varphi}(\varphi_{+},\varphi_{-})
+V_{\Phi,\varphi}(\Phi,\varphi_{+},\varphi_{-})\Bigr]$,
from $G'$ down to $SU(3)_{C} \otimes SU(2)_{L} \otimes U(1)_{Y}$.
Last, the term $V_{\Phi,\varphi,\rho}(\Phi,\varphi_{+},\varphi_{-},\rho)$
makes it possible to break the symmetry down to 
$SU(3)_{C} \otimes U(1)_{Q}$. The detailed formulae for the various
contributions to (3.1) are a bit 
lengthy and can be found in Ref. [3]. 
However, we can describe a few relevant points. On denoting by
$(\Phi \Phi)_{d}$ the $d$-dimensional representation in the tensor
product $210 \otimes 210$, $V_{\Phi}$ is a linear combination of the
norms of $(\Phi \Phi)_{45}, (\Phi \Phi)_{54}, (\Phi \Phi)_{210}$ 
and of the singlet for $(\Phi \Phi)_{210} \times \Phi$. With a
similar notation for $(\varphi \varphi), (\Phi \varphi)$ and
$(\rho \rho)$ terms, one finds that 
$V_{\varphi}(\varphi_{+},\varphi_{-})$ is obtained from the norms
of $(\varphi_{+}\varphi_{+})_{4125}, (\varphi_{-}\varphi_{-})_{4125},
(\varphi_{+}\varphi_{+})_{1050}, (\varphi_{+}\varphi_{+})_{54}$
and $(\varphi_{-}\varphi_{-})_{54}$. The contribution 
$V_{\Phi,\varphi}(\Phi,\varphi_{+},\varphi_{-})$ involves many
cross-terms. However, on choosing the
$SU(3)_{C} \otimes SU(2)_{L} \otimes U(1)_{Y}$ direction, it only
involves $(\Phi \varphi_{+})_{126}, \varphi_{+}$, and the singlet
of $(\varphi_{+}\varphi_{-})_{45} \times (\Phi \Phi)_{45}$.
Last, denoting by $P_{10}$ the most general third-order polynomial
which transforms as a 10-dimensional irreducible representation,
the contribution 
$V_{\Phi,\varphi,\rho}(\Phi,\varphi_{+},\varphi_{-},\rho)$ involves
the singlet of $P_{10} \times \rho$ and the singlet of the
tensor product
$$
\Bigr(q_{1}(\Phi \Phi)_{54}+q_{2}(\varphi_{+}\varphi_{+})_{54}
+q_{2}^{*}(\varphi_{-}\varphi_{-})_{54}\Bigr) \otimes
(\rho \rho)_{54} \; .
$$
\vskip 1cm
\leftline {\bf 4. One-loop effective potential in de Sitter space}
\vskip 1cm
\noindent
Within the framework of inflationary cosmology, the quantization
of non-Abelian gauge fields has been recently studied in the 
case of SU(5) GUT theories [6--8]. In this case one starts from
a bare, Euclidean-time Lagrangian
$$
L={1\over 4}{\mbox {Tr}}({\bf F}_{\mu \nu}{\bf F}^{\mu \nu})
+{1\over 2}{\mbox {Tr}}(D_{\mu}\varphi)(D^{\mu}\varphi)
+V_{0}(\varphi) \; ,
\eqno (4.1)
$$
where both the gauge potential, $A^{\mu}$, and the Higgs
scalar field, $\varphi$, are in the adjoint representation 
of $SU(5)$. The background four-geometry is de Sitter space 
with $S^{4}$ topology. The background-field method is then
used, jointly with the gauge-averaging term first proposed
by 't Hooft
$$
L_{\mbox {g.a.}}={1\over 2}{\widetilde \alpha} \;
{\mbox {Tr}}\Bigr(\nabla_{\mu}A^{\mu}-i{\cal G} \; 
{\widetilde \alpha}^{-1}[\varphi_{0},\varphi]\Bigr)^{2} \; .
\eqno (4.2)
$$
This particular choice is necessary to eliminate from the 
total action cross-terms involving ${\mbox Tr}(\nabla_{\mu}A^{\mu})$
and the commutator $[\varphi_{0},\varphi]$, where
$\varphi_{0}$ is a constant background Higgs field. We now bear in
mind that, by virtue of the Coleman-Weinberg mechanism [9], only
gauge-field loop diagrams contribute to the symmetry-breaking
pattern in the early universe. Thus, after sending 
${\widetilde \alpha} \rightarrow \infty$ (Landau condition), 
and denoting by $\Omega={8\over 3}\pi^{2}a^{4}$ the volume of a
four-sphere of radius $a$, the resulting one-loop effective
potential is [6]
$$
V^{(1)}(\varphi_{0}) \sim V_{0}(\varphi_{0})+{1\over 2\Omega}
{\mbox {log}} \; {\mbox {det}} \; \mu^{-2}
\Bigr[\delta_{ab}(-g_{\sigma \tau}\Box+R_{\sigma \tau})
+g_{\sigma \tau}M_{ab}^{2}(\varphi_{0})\Bigr] \; ,
\eqno (4.3)
$$
since the ghost determinant cancels the longitudinal one.
Denoting by $\psi$ the logarithmic derivative of the $\Gamma$
function, and defining the functions $\cal A$ and $P$ 
by means of
$$
{\cal A}(z) \equiv {z^{2}\over 4}+{z\over 3}
- \left[\int_{2}^{{3\over 2}+\sqrt{{1\over 4}-z}}
+ \int_{1}^{{3\over 2}-\sqrt{{1\over 4}-z}} \right]
y \Bigr(y-{3\over 2}\Bigr)(y-3)\psi(y)dy \; , 
\eqno (4.4)
$$
$$
P(z) \equiv {z^{2}\over 4}+z \; ,
\eqno (4.5)
$$
one thus finds for the $SU(5)$ model [6]
$$
V^{(1)}(\varphi_{0}) \sim V_{0}(\varphi_{0})
-{1\over 2\Omega}\sum_{l=1}^{24}\Bigr[{\cal A}(a^{2}m_{l}^{2})
+P(a^{2}m_{l}^{2}){\mbox {log}}(\mu^{2}a^{2})\Bigr] \; ,
\eqno (4.6)
$$
where the $m_{l}^{2}$ are the 24 eigenvalues of the mass matrix
$M_{ab}^{2}$. 

In the case of $SO(10)$ GUT models, the same method used for 
$SU(5)$ shows that the one-loop effective potential, $V^{(1)}$,
takes the form
$$
V^{(1)} \sim {\widehat V}_{c}-{1\over 2\Omega}\sum_{i=1}^{45}
\Bigr[{\cal A}(a^{2}m_{i}^{2})+P(a^{2}m_{i}^{2})
{\mbox {log}}(\mu^{2}a^{2})\Bigr] \; ,
\eqno (4.7)
$$
where, with the notation of Ref. [4], the renormalizable and 
conformally invariant Higgs potential constructed by using only the 
210-dimensional irreducible representation is
$$
{\widehat V}_{c}=\Bigr({\alpha \over 8}f_{\alpha}
+{\gamma \over 4}f_{\gamma}+{\delta \over 9}f_{\delta}
+(\lambda-\delta)\Bigr){\| \phi_{0} \|}^{4}
+{R\over 12} {\| \phi_{0} \|}^{2} \; ,
\eqno (4.8)
$$
where $R$ denotes the scalar curvature of the background
four-metric. Note that, while (4.7) holds for any irreducible
representation of $SO(10)$, we are only able to evaluate a
{\it particular} form of the one-loop effective potential, 
once the $SU(3) \otimes SU(2) \otimes U(1)$ invariance for the
mass matrix is required to agree with electroweak symmetry [4].
On denoting by $(l,r,x)$ the tensor which behaves as an
$l$-dimensional representation under $SU(3)$, $r$-dimensional
under $SU(2)$, and takes a value $x$ when acted upon by the
$U(1)$ generator, the non-vanishing eigenvalues of the mass
matrix occurring in (4.7) are then found to be
$m_{(1,1,1)}^{2}=m_{(1,1,-1)}^{2}$ with degeneracy $1$,
$m_{(3,1,2/3)}^{2}$ with degeneracy $3$, 
$m_{(3,2,1/6)}^{2}$ with degeneracy $6$, and
$m_{(3,2,-5/6)}^{2}$ also with degeneracy $6$ [4].
\vskip 10cm
\leftline {\bf 5. Results and open problems}
\vskip 1cm
\noindent
The work in Refs. [7, 8] led to a deeper understanding of the
symmetry-breaking pattern for $SU(5)$ models in de Sitter space
first found in Ref. [6], on combining analytic and numerical
techniques with the group-theoretical methods used in Ref. [5].
Although curvature effects modify the flat-space effective potential
by means of the complicated special function defined in (4.4),
the inflationary universe can only slide into either the
$SU(3) \otimes SU(2) \otimes U(1)$ or $SU(4) \otimes U(1)$
extremum [6, 7].

Attention was then focused on $SO(10)$ GUT models for the reasons
described in section 1. On using the particular one-loop effective
potential of section 4, with the mass matrix relevant for the
$SU(3) \otimes SU(2) \otimes U(1)$ symmetry-breaking direction,
it was found in Ref. [4] that, as far as the absolute-minimum
direction is concerned, the flat-space limit of the one-loop
calculation with a de Sitter background does not change the
results relying on the tree-level potential in flat space-time.
Moreover, even when curvature effects are no longer negligible
in the one-loop potential, it was found that the early universe
can only reach the $SU(4)_{PS} \otimes SU(2)_{L} \otimes
SU(2)_{R}$ minimum [4].

However, a constant Higgs field, with de Sitter four-space as
a background in the corresponding one-loop effective potential,
is only a mathematical idealization. A more realistic description
of the early universe is instead obtained on considering a
dynamical space-time such as the one occurring in 
Friedmann-Robertson-Walker models. As a first step in this
programme, electrodynamics for self-interacting scalar fields
in spatially flat Friedmann-Robertson-Walker space-times was
studied in Ref. [10]. In the case of exponentially expanding
universes, the equations for the Bogoliubov coefficients 
describing the coupling of the scalar field to gravity were solved
numerically. They yield a non-local correction to the
Coleman-Weinberg effective potential which does not modify the
pattern of minima found in static de Sitter space. However, such
a correction contains a dissipative term which, accounting for the
decay of the classical configuration in scalar field quanta, may
be relevant for the reheating stage [11--13]. The physical meaning
of the non-local term in the semiclassical field equation was
investigated by evaluating its contribution for various 
background-field configurations [10].

More recently, the work in Ref. [11] has studied the flat-space
limit of the one-loop effective potential for $SO(10)$ GUT
models in spatially flat Friedmann-Robertson-Walker cosmologies.
The numerical integration of the corresponding field equations
shows that a sufficiently long inflationary stage is obtained
for suitable choices of the initial conditions. However, a large
$e$-fold number is only achieved by means of a severe fine
tuning of these initial conditions. Moreover, in the direction
with residual symmetry $SU(4)_{PS} \otimes SU(2)_{L}
\otimes SU(2)_{R}$, one eventually finds parametric resonance [12, 13]
for suitable choices of the free parameters of the tree-level
potential. This leads in turn to the end of inflation [11].

At least two exciting problems are now in sight:
\vskip 0.3cm
\noindent
(i) Can one use the complete form of the tree-level potential 
outlined in section 3 to get a better understanding of the
reheating stage in the early universe ? 
Indeed, the various cross-terms
occurring in (3.1) are negligible during the inflationary era,
but they are important when the second stage of reheating
[12, 13] is considered, since they can mediate the decay of the
massive Higgs into lighter particles [11, 14].
\vskip 0.3cm
\noindent
(ii) How to extend the analysis of non-local effects performed
in Ref. [10] to $SO(10)$ GUT models in 
Friedmann-Robertson-Walker cosmologies ? Can such an investigation
improve the current understanding of dissipative and 
non-dissipative effects [10, 15] in the early universe ?

Although our presentation is far from being exhaustive, the
results and open problems seem to add evidence in favour of a
new age being in sight in cosmology, particle physics and
quantum field theory in curved space-time. We feel that substantial
progress can only result from a fertile interplay between
particle-physics phenomenology and the powerful methods of
relativistic cosmology and quantum field theory. We leave the
reader with this thought, and we hope he will share our
excitement in the course of studying the fundamental problems
of the physics of the early universe.
\vskip 1cm
\leftline {\bf Acknowledgments}
\vskip 1cm
\noindent
I am much indebted to Franco Buccella, Gennaro Miele, Luigi Rosa
and Pietro Santorelli for scientific collaboration on 
group-theoretical methods in the analysis of symmetry-breaking
effects in the early universe. Without their help, my work in
this field would have been unconceivable. I should also thank
Ofelia Pisanti for providing relevant information about the
tree-level potential of $SO(10)$ GUT models. Last, I am grateful
to the INFN for financial support to attend the Second
International Sakharov Conference on Physics.
\vskip 1cm
\leftline {\bf References}
\begin{description}
\item [1.]
Chang D. , Mohapatra R. N. and Parida M. K. (1984)
{\it Phys. Rev. Lett.} {\bf 52}, 1072.
\item [2.]
Tuan S. F. (1992) {\it Mod. Phys. Lett.} A {\bf 7}, 641.
\item [3.]
Acampora F. , Amelino-Camelia G. , Buccella F. , Pisanti O. ,
Rosa L. and Tuzi T. (1995) {\it Nuovo Cimento} A {\bf 108}, 375.
\item [4.]
Esposito G. , Miele G. and Rosa L. (1994) {\it Class. Quantum Grav.}
{\bf 11}, 2031.
\item [5.]
Buccella F. , Ruegg H. and Savoy C. A. (1980) {\it Nucl. Phys.}
B {\bf 169}, 68.
\item [6.]
Allen B. (1985) {\it Ann. Phys.} {\bf 161}, 152.
\item [7.]
Buccella F. , Esposito G. and Miele G. (1992) 
{\it Class. Quantum Grav.} {\bf 9}, 1499.
\item [8.]
Esposito G. , Miele G. and Rosa L. (1993) {\it Class. Quantum Grav.}
{\bf 10}, 1285.
\item [9.]
Coleman S. and Weinberg E. (1973) {\it Phys. Rev.} D {\bf 7}, 1888.
\item [10.]
Esposito G. , Miele G. , Rosa L. and Santorelli P. (1995)
{\it Class. Quantum Grav.} {\bf 12}, 2995.
\item [11.]
Esposito G. , Miele G. and Santorelli P. (1995) {\it Coleman-Weinberg
SO(10) GUT Theories as Inflationary Models} (GR-QC 9512033).
\item [12.]
Kofman L. , Linde A. and Starobinsky A. A. (1994) 
{\it Phys. Rev. Lett.} {\bf 73}, 3195.
\item [13.]
Kofman L. , Linde A. and Starobinsky A. A. (1996)
{\it Phys. Rev. Lett.} {\bf 76}, 1011.
\item [14.]
Buccella F. , Mangano G. , Masiero A. and Rosa L. (1994)
{\it Phys. Lett.} B {\bf 320}, 313.
\item [15.]
Boyanovski D. , de Vega H. J. , Holman R. , Lee D. S and
Singh A. (1995) {\it Phys. Rev.} D {\bf 51}, 4419.
\end{description}

\end{document}